\documentclass{article}

    \usepackage{graphicx} 
    \usepackage{adjustbox} 
    \usepackage{color} 
    \usepackage{enumerate} 
    \usepackage{geometry} 
    \usepackage{amsmath} 
    \usepackage{amssymb} 
    \usepackage{eurosym} 
    \usepackage[mathletters]{ucs} 
    \usepackage[utf8x]{inputenc} 
    \usepackage{fancyvrb} 
    \usepackage{grffile} 
    \usepackage{hyperref}
    \usepackage{longtable} 
    \usepackage{booktabs}  

    \definecolor{orange}{cmyk}{0,0.4,0.8,0.2}
    \definecolor{darkorange}{rgb}{.71,0.21,0.01}
    \definecolor{darkgreen}{rgb}{.12,.54,.11}
    \definecolor{myteal}{rgb}{.26, .44, .56}
    \definecolor{gray}{gray}{0.45}
    \definecolor{lightgray}{gray}{.95}
    \definecolor{mediumgray}{gray}{.8}
    \definecolor{inputbackground}{rgb}{.95, .95, .85}
    \definecolor{outputbackground}{rgb}{.95, .95, .95}
    \definecolor{traceback}{rgb}{1, .95, .95}
    \definecolor{red}{rgb}{.6,0,0}
    \definecolor{green}{rgb}{0,.65,0}
    \definecolor{brown}{rgb}{0.6,0.6,0}
    \definecolor{blue}{rgb}{0,.145,.698}
    \definecolor{purple}{rgb}{.698,.145,.698}
    \definecolor{cyan}{rgb}{0,.698,.698}
    \definecolor{lightgray}{gray}{0.5}
    
    \definecolor{darkgray}{gray}{0.25}
    \definecolor{lightred}{rgb}{1.0,0.39,0.28}
    \definecolor{lightgreen}{rgb}{0.48,0.99,0.0}
    \definecolor{lightblue}{rgb}{0.53,0.81,0.92}
    \definecolor{lightpurple}{rgb}{0.87,0.63,0.87}
    \definecolor{lightcyan}{rgb}{0.5,1.0,0.83}
    
    \DefineVerbatimEnvironment{Highlighting}{Verbatim}{commandchars=\\\{\}}

\title{Proposed Approximate Dynamic Programming for Pathfinding under Visible Uncertainty}

\author{Bryan A. Knowles and Mustafa Atici}

\makeatletter
\def\PY@reset{\let\PY@it=\relax \let\PY@bf=\relax%
    \let\PY@ul=\relax \let\PY@tc=\relax%
    \let\PY@bc=\relax \let\PY@ff=\relax}
\def\PY@tok#1{\csname PY@tok@#1\endcsname}
\def\PY@toks#1+{\ifx\relax#1\empty\else%
    \PY@tok{#1}\expandafter\PY@toks\fi}
\def\PY@do#1{\PY@bc{\PY@tc{\PY@ul{%
    \PY@it{\PY@bf{\PY@ff{#1}}}}}}}
\def\PY#1#2{\PY@reset\PY@toks#1+\relax+\PY@do{#2}}

\expandafter\def\csname PY@tok@nc\endcsname{\let\PY@bf=\textbf\def\PY@tc##1{\textcolor[rgb]{0.00,0.00,1.00}{##1}}}
\expandafter\def\csname PY@tok@mo\endcsname{\def\PY@tc##1{\textcolor[rgb]{0.40,0.40,0.40}{##1}}}
\expandafter\def\csname PY@tok@k\endcsname{\let\PY@bf=\textbf\def\PY@tc##1{\textcolor[rgb]{0.00,0.50,0.00}{##1}}}
\expandafter\def\csname PY@tok@err\endcsname{\def\PY@bc##1{\setlength{\fboxsep}{0pt}\fcolorbox[rgb]{1.00,0.00,0.00}{1,1,1}{\strut ##1}}}
\expandafter\def\csname PY@tok@gs\endcsname{\let\PY@bf=\textbf}
\expandafter\def\csname PY@tok@sr\endcsname{\def\PY@tc##1{\textcolor[rgb]{0.73,0.40,0.53}{##1}}}
\expandafter\def\csname PY@tok@vi\endcsname{\def\PY@tc##1{\textcolor[rgb]{0.10,0.09,0.49}{##1}}}
\expandafter\def\csname PY@tok@go\endcsname{\def\PY@tc##1{\textcolor[rgb]{0.53,0.53,0.53}{##1}}}
\expandafter\def\csname PY@tok@kn\endcsname{\let\PY@bf=\textbf\def\PY@tc##1{\textcolor[rgb]{0.00,0.50,0.00}{##1}}}
\expandafter\def\csname PY@tok@gt\endcsname{\def\PY@tc##1{\textcolor[rgb]{0.00,0.27,0.87}{##1}}}
\expandafter\def\csname PY@tok@kc\endcsname{\let\PY@bf=\textbf\def\PY@tc##1{\textcolor[rgb]{0.00,0.50,0.00}{##1}}}
\expandafter\def\csname PY@tok@nl\endcsname{\def\PY@tc##1{\textcolor[rgb]{0.63,0.63,0.00}{##1}}}
\expandafter\def\csname PY@tok@kr\endcsname{\let\PY@bf=\textbf\def\PY@tc##1{\textcolor[rgb]{0.00,0.50,0.00}{##1}}}
\expandafter\def\csname PY@tok@mb\endcsname{\def\PY@tc##1{\textcolor[rgb]{0.40,0.40,0.40}{##1}}}
\expandafter\def\csname PY@tok@ss\endcsname{\def\PY@tc##1{\textcolor[rgb]{0.10,0.09,0.49}{##1}}}
\expandafter\def\csname PY@tok@vg\endcsname{\def\PY@tc##1{\textcolor[rgb]{0.10,0.09,0.49}{##1}}}
\expandafter\def\csname PY@tok@sc\endcsname{\def\PY@tc##1{\textcolor[rgb]{0.73,0.13,0.13}{##1}}}
\expandafter\def\csname PY@tok@gd\endcsname{\def\PY@tc##1{\textcolor[rgb]{0.63,0.00,0.00}{##1}}}
\expandafter\def\csname PY@tok@nd\endcsname{\def\PY@tc##1{\textcolor[rgb]{0.67,0.13,1.00}{##1}}}
\expandafter\def\csname PY@tok@si\endcsname{\let\PY@bf=\textbf\def\PY@tc##1{\textcolor[rgb]{0.73,0.40,0.53}{##1}}}
\expandafter\def\csname PY@tok@cp\endcsname{\def\PY@tc##1{\textcolor[rgb]{0.74,0.48,0.00}{##1}}}
\expandafter\def\csname PY@tok@il\endcsname{\def\PY@tc##1{\textcolor[rgb]{0.40,0.40,0.40}{##1}}}
\expandafter\def\csname PY@tok@sh\endcsname{\def\PY@tc##1{\textcolor[rgb]{0.73,0.13,0.13}{##1}}}
\expandafter\def\csname PY@tok@ni\endcsname{\let\PY@bf=\textbf\def\PY@tc##1{\textcolor[rgb]{0.60,0.60,0.60}{##1}}}
\expandafter\def\csname PY@tok@ne\endcsname{\let\PY@bf=\textbf\def\PY@tc##1{\textcolor[rgb]{0.82,0.25,0.23}{##1}}}
\expandafter\def\csname PY@tok@na\endcsname{\def\PY@tc##1{\textcolor[rgb]{0.49,0.56,0.16}{##1}}}
\expandafter\def\csname PY@tok@c\endcsname{\let\PY@it=\textit\def\PY@tc##1{\textcolor[rgb]{0.25,0.50,0.50}{##1}}}
\expandafter\def\csname PY@tok@mi\endcsname{\def\PY@tc##1{\textcolor[rgb]{0.40,0.40,0.40}{##1}}}
\expandafter\def\csname PY@tok@m\endcsname{\def\PY@tc##1{\textcolor[rgb]{0.40,0.40,0.40}{##1}}}
\expandafter\def\csname PY@tok@ge\endcsname{\let\PY@it=\textit}
\expandafter\def\csname PY@tok@nf\endcsname{\def\PY@tc##1{\textcolor[rgb]{0.00,0.00,1.00}{##1}}}
\expandafter\def\csname PY@tok@gh\endcsname{\let\PY@bf=\textbf\def\PY@tc##1{\textcolor[rgb]{0.00,0.00,0.50}{##1}}}
\expandafter\def\csname PY@tok@sb\endcsname{\def\PY@tc##1{\textcolor[rgb]{0.73,0.13,0.13}{##1}}}
\expandafter\def\csname PY@tok@w\endcsname{\def\PY@tc##1{\textcolor[rgb]{0.73,0.73,0.73}{##1}}}
\expandafter\def\csname PY@tok@se\endcsname{\let\PY@bf=\textbf\def\PY@tc##1{\textcolor[rgb]{0.73,0.40,0.13}{##1}}}
\expandafter\def\csname PY@tok@kt\endcsname{\def\PY@tc##1{\textcolor[rgb]{0.69,0.00,0.25}{##1}}}
\expandafter\def\csname PY@tok@mf\endcsname{\def\PY@tc##1{\textcolor[rgb]{0.40,0.40,0.40}{##1}}}
\expandafter\def\csname PY@tok@vc\endcsname{\def\PY@tc##1{\textcolor[rgb]{0.10,0.09,0.49}{##1}}}
\expandafter\def\csname PY@tok@cm\endcsname{\let\PY@it=\textit\def\PY@tc##1{\textcolor[rgb]{0.25,0.50,0.50}{##1}}}
\expandafter\def\csname PY@tok@nv\endcsname{\def\PY@tc##1{\textcolor[rgb]{0.10,0.09,0.49}{##1}}}
\expandafter\def\csname PY@tok@nt\endcsname{\let\PY@bf=\textbf\def\PY@tc##1{\textcolor[rgb]{0.00,0.50,0.00}{##1}}}
\expandafter\def\csname PY@tok@gu\endcsname{\let\PY@bf=\textbf\def\PY@tc##1{\textcolor[rgb]{0.50,0.00,0.50}{##1}}}
\expandafter\def\csname PY@tok@bp\endcsname{\def\PY@tc##1{\textcolor[rgb]{0.00,0.50,0.00}{##1}}}
\expandafter\def\csname PY@tok@sd\endcsname{\let\PY@it=\textit\def\PY@tc##1{\textcolor[rgb]{0.73,0.13,0.13}{##1}}}
\expandafter\def\csname PY@tok@gr\endcsname{\def\PY@tc##1{\textcolor[rgb]{1.00,0.00,0.00}{##1}}}
\expandafter\def\csname PY@tok@s\endcsname{\def\PY@tc##1{\textcolor[rgb]{0.73,0.13,0.13}{##1}}}
\expandafter\def\csname PY@tok@no\endcsname{\def\PY@tc##1{\textcolor[rgb]{0.53,0.00,0.00}{##1}}}
\expandafter\def\csname PY@tok@gp\endcsname{\let\PY@bf=\textbf\def\PY@tc##1{\textcolor[rgb]{0.00,0.00,0.50}{##1}}}
\expandafter\def\csname PY@tok@cs\endcsname{\let\PY@it=\textit\def\PY@tc##1{\textcolor[rgb]{0.25,0.50,0.50}{##1}}}
\expandafter\def\csname PY@tok@nn\endcsname{\let\PY@bf=\textbf\def\PY@tc##1{\textcolor[rgb]{0.00,0.00,1.00}{##1}}}
\expandafter\def\csname PY@tok@ow\endcsname{\let\PY@bf=\textbf\def\PY@tc##1{\textcolor[rgb]{0.67,0.13,1.00}{##1}}}
\expandafter\def\csname PY@tok@kd\endcsname{\let\PY@bf=\textbf\def\PY@tc##1{\textcolor[rgb]{0.00,0.50,0.00}{##1}}}
\expandafter\def\csname PY@tok@mh\endcsname{\def\PY@tc##1{\textcolor[rgb]{0.40,0.40,0.40}{##1}}}
\expandafter\def\csname PY@tok@o\endcsname{\def\PY@tc##1{\textcolor[rgb]{0.40,0.40,0.40}{##1}}}
\expandafter\def\csname PY@tok@s2\endcsname{\def\PY@tc##1{\textcolor[rgb]{0.73,0.13,0.13}{##1}}}
\expandafter\def\csname PY@tok@c1\endcsname{\let\PY@it=\textit\def\PY@tc##1{\textcolor[rgb]{0.25,0.50,0.50}{##1}}}
\expandafter\def\csname PY@tok@kp\endcsname{\def\PY@tc##1{\textcolor[rgb]{0.00,0.50,0.00}{##1}}}
\expandafter\def\csname PY@tok@gi\endcsname{\def\PY@tc##1{\textcolor[rgb]{0.00,0.63,0.00}{##1}}}
\expandafter\def\csname PY@tok@sx\endcsname{\def\PY@tc##1{\textcolor[rgb]{0.00,0.50,0.00}{##1}}}
\expandafter\def\csname PY@tok@nb\endcsname{\def\PY@tc##1{\textcolor[rgb]{0.00,0.50,0.00}{##1}}}
\expandafter\def\csname PY@tok@s1\endcsname{\def\PY@tc##1{\textcolor[rgb]{0.73,0.13,0.13}{##1}}}

\makeatother

    \definecolor{incolor}{rgb}{0.0, 0.0, 0.5}
    \definecolor{outcolor}{rgb}{0.545, 0.0, 0.0}

    \hypersetup{
      breaklinks=true,  
      colorlinks=true,
      urlcolor=blue,
      linkcolor=darkorange,
      citecolor=darkgreen,
      }
    
\begin{document}

    \maketitle

    \textbf{Abstract}

    Continuing our preleminary work \cite{knowles14}, we define the
safest-with-sight pathfinding problems and explore its solution using
techniques borrowed from measure-theoretic probability theory. We find a
simple recursive definition for the probability that an ideal pathfinder
will select an edge in a given scenario of an uncertain network where
edges have probabilities of failure and vertices provide ``vision'' of
edges via lines-of-sight. We propose an approximate solution based on
our theoretical findings that would borrow techniques from approximate
dynamic programming.

    \section{Introduction}\label{introduction}

    We introduce a probabilistic-decision variant of the classic pathfinding
problem defined on a directed acylic graph \cite{wilson80} where each
edge has some probability of ``failure'' and each vertex has ``vision''
of a set of edges. That is, once the pathfinder has reached a vertex, it
can ``know'' whether the edges within the sight of that vertex are up or
down; the status of these edges are said to not change during the
duration of a single ``trial.'' How, then, should the pathfinder behave
if it wishes to take the ``safest'' (most likely to succeed) path taking
this ``sight'' into account?

Although this safest-with-sight problem, as we will refer to it
throughout, is simple to state, and we have restricted it to directed
acyclic graphs, which generally reduce the complexity of problems, we
believe that the introduction of vision into the mix makes this problem
computationally hard.

Naturally, this being a graph-theory-grounded problem, we wish to
determine a greedy algorithm \cite{wilson80} to answer the query,
``given a current scenario, will an \emph{ideal} pathfinder's \emph{one
next} move be $x$?'' However, this being a problem steeped in
uncertainties, we must also accomplish this task probabilistically,
defining a decision function by borrowing techniques from
measure-theoretic probability theory \cite{pollard01}. So, to calculate
the solution in the general case, where we find that a greedy solution
cannot work, we find a dynamic algorithm and propose, to reduce the
computational complexities in that unmodified algorithm, to use an
approximate dynamic scheme instead \cite{powell11}.

In section 2 we briefly discuss a portion of the literature on the
subject, in section 3 we discuss our laying the groundwork for a
theoretical solution to this problem, and in section 4 we conclude by
stating our goals in implementing an approximate solution for the
general case that is exact and polynomial in certain cases of the
problem.

    \section{Previous Work}\label{previous-work}

    We have been unable to find any publication on the same problem as ours
or on a problem that is immediately reducible to safest-with-sight. The
key difference between our definition and others is that of vision--the
ability of the pathfinder to select options that will make future
options more informed. It is by this distinction that we mean ``visible
uncertainty.''

However, the literature does contain works related to \emph{components}
of ours. In a preliminary article \cite{knowles14}, we explore several
works on undirected, random, and directed-acyclic graphs each with
edge-risk probabilities, and others on paths through and relationships
between layers of probabilistic networks.

Having quickly reached problems with approaching safest-with-sight from
a purely graph-theoretical perspective after releasing our preliminary
work in preprint, we looked further into uncertainty and probability
theory itself.

In a now classic work, Bart Kosko introduces fuzzy logic, now a staple
in the machine learning literature, which does away with the binary of
true and false in favor of a model based on partial membership within
sets \cite{kosko93}. David Pollard writes on measure-theoretic
probability theory, which reevaluates how we construct probabilities as
not densities, but as measure functions or expected values of inclusion
within a set of outcomes \cite{pollard01}. And in a collection of works
by Springer, the foundations of applying fuzzy logic and uncertainty is
set forth \cite{springer10}.

None of these publications has been more influential on our research
more than Pollard's, as it forced us to work axiomatically from set
theoretical definitions, making clear the exact influence that vision
has on ``the math.'' It is in the following statement, and no more, that
vision has its priciple effects: I will never (probability equals zero)
decide to take an edge that is both down and in my accumulated
line-of-sight.

    \section{Theoretical Foundations}\label{theoretical-foundations}

    \textbf{Problem.} An instance of the \emph{safest-with-sight} problem is
defined by the tuple $(G, \beta, t_{sd})$. First, $G=(V, E, W)$, where
$V$ is a set of vertices, $E$ is a set of edges and $ij \in E$ implies
an edge exists between $i$ and $j$ such that $i < j$, and $W$ is a set
of lines-of-sight and $ijk \in W$ implies vertex $i$ has line-of-sight
to edge $jk$ such that $i \le j$. Second, $\beta$ is a set of parameters
where $\beta_{\alpha ij}$ is the probability of event $\alpha_{ij}$, in
which edge $ij$ is down or obstructed. Finally, we are given the task
$t_{sd}$, representing the starting vertex $s$ and destination vertex
$d$. Given a possible first step $si$ that the pathfinder could take,
assuming the pathfinder behaves ideally for the following properties,
decide whether the pathfinder will take that first step:

\begin{enumerate}
\def\labelenumi{\arabic{enumi}.}
\itemsep1pt\parskip0pt\parsep0pt
\item
  if the pathfinder crosses an edge that is down, it can no longer move
  and fails the trial immediately
\item
  the pathfinder can only follow directed, simple paths that start at
  the starting point and end either at the destination or the first
  visited point at which the pathfinder knows that all remaining paths
  end in ``dead-ends''
\item
  the pathfinder will never attempt to cross edges that it knows are
  down or lead only to dead-ends
\item
  the pathfinder always selects the path that maximizes its probability
  of successfully reaching the destination with respects to its current
  knowledge
\item
  the pathfinder always knows the layout of the graph, including
  lines-of-sight
\item
  the pathfinder only knows the up/down status of edges that are in a
  line-of-sight of any vertex it has ever visited during the current
  trial
\item
  extraneous edges have been removed, so any unobstructed path the
  pathfinder takes will lead it to the destination
\item
  the pathfinder stops at each vertex to consider new information from
  new lines-of-sight, and it may use this information to reroute its
  current course such that the probability of success remains optimized
  with respects to the pathfinder's current knowledge
\item
  depending on the application, a definition of ``tie-broken'' is given
  that imposes a total ordering on the edges without respect to
  probabilities of success; if no definition for tiebreaking can be
  given, we suggest using the outward indices of edges for tiebreaking
  so that the pathfinder will select the optimal edge with the highest
  outward index
\end{enumerate}

\textbf{Theorem 1.} The safest-with-sight problem cannot be solved with
a greedy algorithm.

\textbf{Proof.} The probability of success of a ``first edge'' is
determined, in part, by the probability of future edges and their
probabilities of being taken; however, the probabilities of a future
edge is likewise determined by the probabilities of first edges, since
they determine what set of lines-of-sight may be available once that
future edge has been reached. In other words, the behavior of the
pathfinder at the first step and future steps are caught in a
chicken-and-egg problem. This prevents safest-with-sight from meeting
the greedy-choice property, since solutions to problems depend on
solutions to subproblems \cite{cormen90}. Therefore, safest-with-sight
cannot be greedy. \emph{q.e.d.}

\textbf{Definition.} We define a function $\Phi$ that digitizes
statements as $1$s or $0$s.

\begin{align}
    \Phi(x) &= \begin{cases}
        1 & x \text{ is true} \\
        0 & \text{Otherwise}
    \end{cases}\\
    \text{if } x_i \text{ are independant} &\text{ then } \forall_i \Phi(x_i) = \prod_i \Phi(x_i)
\end{align}

\textbf{Definition.} We define $P$ as an underlying a probability
measure for solving this problem, choosing notation such that $P$ maps
\emph{queries encoded as a summed series of terms} to the space
${\bf B}$ of basis $B$. In-depth measure-theoretic details do not matter
in our application, as we will replace all occurrences of $P$ in our
problem with either event methods, which correspond with computable
functions, or with conditional statements of constant values.

First, our outcome space is defined as follows, letting each $\omega$
representing a vector or sum of terms where $t$ is a term representing
the task of the trial, each $a$ is a term whether some edge was up or
down, each $d$ is a term describing whether some edge was taken or not
taken by the pathfinder, and each $g$ is a term describing the graph's
edges, vertices, and lines-of-sight.

\begin{align}
    \Omega &= \left\{ \omega \mid \omega = t + \sum a + \sum d + \sum g \right\}
\end{align}

Next, $P$, ${\bf F}$, and ${\bf B}$ are defined as follows.

\begin{align}
    P: {\bf F} &\to {\bf B} \\
    \Omega &\in {\bf F} \\
    \text{if } \omega \in \Omega &\text{ then } \{ \omega \} \in {\bf F} \\
    \text{if } A \in {\bf F} &\text{ then } \Omega \backslash A \in {\bf F} \\
    \text{if } A_1, A_2, \dots \in {\bf F} &\text{ then } \bigcap_k A_k \in {\bf F}\\
    \beta \cup \left\{ 0 \right\} &\subseteq {\bf B} \\
    \text{if } b \in {\bf B} &\text{ then } 1-b \in {\bf B} \\
    \text{if } b_1, b_2, \dots \in {\bf B} &\text{ then } \prod_k b_k \in {\bf B}
\end{align}

And we state the following about $P$ as it is used to encode queries
about the pathfinder's behavior:

\begin{align}
    P(0) &= 1 \\
    P(a_1 + a_2 + \dots) &= P(a_1 \text{ and } a_2 \text{ and } \dots) \\
    P(\bar a) &= 1 - P(a) \\
    P(a + \bar a) &= 0 \\
    \text{if } a_i \text{ are independant} \text{ then } \exists_i P(a_i) &= 1 - \prod_i \left[ 1 - P(a_i) \right] \\
    \exists_{i} \left[ P(a_i + \xi_1 \mid b_i + \xi_2) \right] &= { \sum_{i} \left[ P(a_i + \sum_{j=1}^{i-1} \bar a_j + b_i + \sum_{j=1}^{i-1} \bar b_j + \xi_1 + \xi_2) \right] \over \sum_{i} \left[ b_i + \sum_{j=1}^{i-1} \bar b_j + \xi_2) \right] } \\
    P(a) &= \beta_a,\, \forall_{a \in \text{ parametric terms}} \\
    P(w+\dots) &= 0,\, \forall_{w \in \text{ contradictions}} \\
\end{align}

\begin{align}
    \text{if } i < s \text{ or } j > d& \text{ then } P(t_{sd} + a_{ij} + \dots) = P(t_{sd} + \dots) \\
    \text{ev}(a) &= \sum_{a'} P(a' \mid a) P(\text{ev} \mid a') \\
    \text{iff } a \text{ is M.R. on ev}& \text{ then } \text{ev}(a) = P(\text{ev} \mid a)
\end{align}

By the term M.R. above, we mean ``maximally restrictive.'' That is, the
event method ${\tt ev}$, given $a$, has no change in value for any $b$
disjoint from $a$ in ${\tt ev}(a+b)$. Therefore, $a$ contains as much
information as possible for determining ${\tt ev}$'s value.

\textbf{Definition.} We define the set of terms used to encode events
as: $t_{sd}$, the task term; $\alpha_{ij}$, terms for the event that
edge $ij$ is down; $s_{ijk}$, terms for the event that vertex $i$ has
vision of edge $jk$; and $\delta_{ij}$, terms for the event that the
pathfinder \emph{chose} to travel along edge $ij$, but not necessarily
traveled along it safely.

\textbf{Definition.} We complete the definition of $P$ by defining what
we mean above by $\tt{parametric terms}$ and $\tt{contradictions}$:

\begin{itemize}
\itemsep1pt\parskip0pt\parsep0pt
\item
  the parametric terms are those for which we are given a
  $\beta$-parameter, such as $\alpha_{ij}$ in $\beta_{\alpha ij}$
\item
  if $a$ and $b$ are parametric terms, then $P(a \mid b) = P(a)$ unless
  $b=\bar a$, in which case $P(a \mid b) = 0$
\item
  the contradictions are the minimum set of expression which, for all
  queries containing one or more of those expressions, the probability
  must be zero
\end{itemize}

This set of contradictions corresponds to exactly the following, derived
directly from the properties of the problem and definition of $P$:

\begin{itemize}
\itemsep1pt\parskip0pt\parsep0pt
\item
  classic--appealing to $(15)$, $P(a + \bar a) = 0$
\item
  restriction--appealing to property 1,
  $P(\delta_{ij} + \delta_{jk} + \alpha_{ij}) = 0$
\item
  simplicity--appealing to property 2,
  $P(\delta_{ij} + \delta_{ik}) = 0$ and
  $P(t_{sd} + \delta_{jk} \sum_{ij} \bar \delta_{ij}) = 0 \text{ if } j \neq s$
\item
  refusal--appealing to property 3,
  $P(\delta_{ij} + \delta_{km} + \alpha_{km} + s_{ikm}) = 0$
\item
  dead-ends--appealing to properties 2 and 3,
  $P(\delta_{ij} \mid t_{sd} + \dots) = 0 \text{ if } j \neq d \text{ and } \forall_{jk} \left[ P(\delta_{jk} \mid t_{sd} + \dots) = 0 \right]$
\item
  suboptimal--appealing to property 4,
  $P(\delta_{si} \mid t_{sd} + \dots) = \text{optimal}(\delta_{si} + t_{sd} + \dots) \times \text{tiebroken}(\delta_{si} + t_{sd} + \dots)$
\end{itemize}

\textbf{Definition.} We define a set of event methods
${\tt select}_{si}$ for all edges $si$ that, on maximally restrictive
input, equals $1$ only where the pathfinder would select edge $si$ given
the current task $t_{sd}$, knowledge $\xi$, and lines-of-sight $S$;
otherwise, it equals $0$. It is important to note that this event method
is simply the functional equivalent of $\delta_{si}$, only given an
easier to express name. This definition requires the definitions of
event methods ${\tt optimal}_{si}$, ${\tt tiebroken}_{si}$, and
${\tt success}_{si}$, as well as $S'$, which corresponds to the
lines-of-sight of a subproblem where the vision provided by $S$ has been
``copied'' to all other vertices and truncated such that no vertex
``sees behind itself.''

\begin{align*}
    &\text{select}_{si}(t_{sd} + S + \xi) =
        \text{optimal}_{si}(t_{sd} + S + \xi) \times
        \text{tiebroken}_{si}(t_{sd} + S + \xi) \\
    &\text{where optimal}_{si}(t_{sd} + S + \xi) = \forall_{sj} \Phi \left[
        \text{success}_{si}(t_{sd} + S + \xi) \ge
        \text{success}_{sj}(t_{sd} + S + \xi)
    \right] \\
    &\text{and success}_{sd}(t_{sd} + S + \xi) =
        P(\bar \alpha_{sd} \mid \delta_{sd} + t_{sd} + S + \xi) \\
    &\text{and success}_{si}(t_{sd} + S + \xi) =
        P(\bar \alpha_{si} \mid \delta_{si} + t_{sd} + S + \xi) \times
        \sum_{\xi'} \left[ P(\xi' \mid \xi) \times \max_{ik} \text{success}_{ik}(t_{id} + S' + \xi') \right]
\end{align*}

\textbf{Lemma 1.} If ${\tt success}_{si}$ is correctly the probability
that the pathfinder will reach the destination (given only what it could
know at $s$ and assuming it has decided to traverse $si$) then
${\tt select}_{si}$ correctly determines whether the pathfinder will
choose to traverse $si$.

\textbf{Proof.} Appealing to properties 3 and 9, the pathfinder will
select an edge iff it is optimal and tiebroken. Obviously too, an edge
is optimal iff there exists no other available edge that, according to
the pathfinder's current knowledge, has a higher probability of success.

We've defined ${\tt optimal}_{si}$ as a digitization of this notion, so
when the input is maximally restrictive, this event method maps to
$\{0, 1\}$. If we assume the input of ${\tt select}_{si}$ is maximally
restrictive on both ${\tt select}_{si}$ and ${\tt optimal}_{si}$, then,
since the same input is passed to both, ${\tt select}_{si}$ will also
map to $\{0, 1\}$, being itself the product of two such mappings. We may
assume that the initial input, given by the problem, is maximally
restrictive in this way, since it corresponds to precisely the knowledge
held by the pathfinder while at $s$, the start of the trial. Because no
other input is given to ${\tt select}_{si}$, these assumptions about
maximally restrictive input hold for ${\tt select}_{si}$ and
${\tt optimal}_{si}$ always. That is, there is no need to worry about
weighted sums, as defined in $(21)$ and $(22)$.

Therefore, if ${\tt success}_{si}$ is correct, then ${\tt optimal}_{si}$
will yield the proper value and, in turn, ${\tt select}_{si}$ as well.
\emph{q.e.d.}

\textbf{Lemma 2.} Assuming the pathfinder has decided to traverse edge
$si$, its probability of success (according only to the pathfinder's
knowledge) is the probability it successfully crosses $si$ and it
succeeds from some subproblem where its starting point is instead $i$
and lines-of-sight everywhere have been modifed to include the
lines-of-sight provided by $s$ in the original problem.

\textbf{Proof.} When the pathfinder is traversing an edge, it can fail
immediately. In the event that it does not, it, appealing to property 8,
stops to consider new information. If $s$ provided no lines-of-sight
that were not provided by $i$, then its probability of success after
crossing $si$ is no different than its overall probability if it were to
begin from $i$ in the first place. If $s$ \emph{did} provide
lines-of-sight useful after $i$ that were not provided by $i$
originally, because, appealing to property 6, the state of the network
does not change during a single trial and the pathfinder does not forget
information, we can model its ``remembering'' by copying the vision
provided by $s$ to all vertices ahead of it, adding no information to
the subproblem that would not have already been known anyway. Therefore,
by appealing to the properties of the problem, we can derive this lemma
directly. \emph{q.e.d.}

\textbf{Theorem 2.} ${\tt success}_{si}$ is correct as required by lemma
1.

\textbf{Proof.} First, consider what maximally restritive input to
${\tt success}_{si}$ might look like: the maximum information that the
pathfinder could have that could change the probability of its success,
according only to the knowledge that it could have obtained via
lines-of-sight, is where $\xi$ contains $\alpha_{ij}$ or
$\bar \alpha_{ij}$ terms for all edges $ij$ referenced in $S$. This is
precisely the sort of input assumed in lemma 1 to be passed to
${\tt success}_{si}$ by ${\tt optimal}_{si}$.

Next, appealing to lemma 2 and $(21)$, we write ${\tt success}_{si}$ as
the following:

\begin{align}
    \text{success}_{si}(t_{sd} + S + \xi) =
        &\sum_{\xi'} \left[
            P(\xi' \mid \xi) \times P(\bar \alpha_{si} \mid \delta_{si} + t_{sd} + S + \xi')
        \right] \times \\
        &\sum_{\xi'} \sum_{ik} \left[
            P(\xi' \mid \xi) \times 
            \text{select}_{ik}(t_{id} + S' + \xi') \times
            \text{success}_{ik}(t_{id} + S' + \xi')
        \right]
\end{align}

Consider how the probability that an edge is up is only affected by
whether the edge is already known to be up/down (referenced in $S$ and
$\alpha$ or $\bar \alpha$ is in $\xi$). Hence, the only properties of
the problem that have an effect on $P(\bar \alpha_{si} \mid \dots)$ are
$(15)$ and the refusal contradiction; these require only terms that
would already be in a maximally restrictive input to
${\tt success}_{si}$. Therefore, for that the first weighted sum we can
appeal to $(23)$, at least only when the input to ${\tt success}_{si}$
is given to be maximally restrictive.

${\tt success}_{si}$ is only ever given maximally restrictive input in
our definitions, since ${\tt optimal}_{si}$ passes it input that is
optimally restrictive and when ${\tt success}_{si}$ invokes
${\tt success}_{ik}$, it does so only under a weighted sum iterating
over all maximally restrictive inputs to the subproblem that contain
$\xi$. Therefore, we can appeal to $(23)$ for all cases covered by our
definitions and rewrite the first weighted sum as just
$P(\bar \alpha_{si} \mid \dots)$.

For the second weighted sum, note that $\xi$ is not guaranteed to be
maximally restrictive for the subproblem, since $S'$ might reference an
edge $km$ such that neither $\alpha_{km}$ nor $\bar \alpha_{km}$ are in
$\xi$; therefore, the iteration over $\xi'$ must be performed and cannot
be simplified away as in the first weighted sum.

However, consider how the inner sum operates: because
${\tt select}_{si}$ maps to $\{0, 1\}$, and to $1$ only where optimal
and tiebroken, we can write this inner sum simpler as the maximum
subproblem, where the notion of maximizing is with respects to both
${\tt optimal}_{ik}$ and ${\tt tiebroken}_{ik}$.

Taking all of these simplifications into account, we produce exactly the
recursive definition of ${\tt success}_{si}$ from above. With this in
place, the base case definition is trivial to show: once the pathfinder
has decided to traverse some final edge leading directly from $s$ into
$d$, its probability of success is simply that of whether or not it
successfully that edge--there is no ``succeeds later'' to worry about.
Therefore, the recursive definition is just
$P(\bar \alpha_{si} \mid \dots)$, what remains after simplifying the
first weighted sum and removing the (in the base case) unnecessary
second weighted sum.

Finally, since the first invocation of ${\tt success}_{si}$ is given the
proper inputs, it is correct under lemma 2, it passes the correct inputs
to future invocations, and the base case is also correct under lemma 2,
${\tt success}_{si}$ must be correct in determing the probability of
success as required by lemma 1. \emph{q.e.d.}

    \section{Conclusions and Future Work}\label{conclusions-and-future-work}

    The solution we have defined to the safest-with-sight problem, a
decision problem based on $\tt{success}_{si}$, is not computationally
efficient. When each path through to a certain node would produce a
different set of known edges, via accumulated lines-of-sight, then an
exact algorithm would need to calculate each of those paths, a
PCOUNT-time solution. In cases where the set of known edges at a node is
always the same, then solutions to recursive cases of
$\tt{success}_{si}$ could be cached using dynamic programming techniques
and the runtime reduced to O(E), since each edge would only need to be
visited once by the algorithm, as in the following two examples:

\begin{itemize}
\itemsep1pt\parskip0pt\parsep0pt
\item
  no lines-of-sight exist, so vision is always empty
\item
  lines-of-sight are only of immediate neighbors, so vision cannot
  accumulate
\end{itemize}

With the cache-based improvement in mind, we propose an approximate
solution to the general case of safest-with-sight, where solutions to
recursive cases of $\tt{score}$ are cached and, when a
\emph{similar-enough} case has already been cached for the same edge,
that cached solution would be used in hopes that the scores will not
differ greatly. This approach will have to be done with care to find a
balance between space required to cache results and time required to
produce accurate solutions \cite{powell11}.

    \section{Special Thanks}\label{special-thanks}

    Special thanks to IPython/Jupyter \cite{ipython} for providing a free
and open source platform used extensively when developing this work.


\bibliographystyle{plain}
\bibliography{references}

    \end{document}